\documentclass{ws-as3n}

\def\beq{\begin{equation}}
\def\eeq{\end{equation}}
\def\beqa{\begin{eqnarray}}
\def\eeqa{\end{eqnarray}}

\begin{document}
\thispagestyle{empty}

\begin{flushright}
Mar 2000
\end{flushright}

\vspace*{.5in}

\begin{center}
{\bf  \Large Could a Slepton be the Zee boson?}$^\star$\\
\vspace*{.5in}
{\bf Kingman Cheung$^1$ and Otto C.W. Kong$^2$ }\\[.05in]
$^1${\it Department of Physics, University of California, Davis, CA 95616 USA}\\[.05in]
$^2${\it Institute of Physics, Academia Sinica, Nankang, Taipei, TAIWAN 11529}

\vspace*{.8in}
{Abstract}\\
\end{center}
We study the feasibility of incorporating
the popular Zee model for neutrino mass in the 
framework of the supersymmetric standard model. 
While a singlet slepton has the right quantum number
to play the role of the Zee charged scalar boson, 
the SUSY framework introduces extra contributions. 
We derive conditions for the Zee contributions to 
dominate, hence retaining the flavor of the Zee 
model.

\noindent

\vfill
\noindent --------------- \\
$^\star$ Talk presented by O.K. at PASCOS 99 
(Dec 10-16), Lake Tahoe, USA\\
 --- submission for the proceedings.  
 
\clearpage
\addtocounter{page}{-1}

\title{Could a Slepton be the Zee boson?}

\author{Kingman Cheung }

\address{Department of Physics, University of California, Davis, 
CA 95616 USA \\E-mail: cheung@gluon.ucdavis.edu}

\author{Otto C.W. Kong}

\address{Institute of Physics, Academia Sinica, Nankang, Taipei, TAIWAN 11529\\
E-mail: kongcw@phys.sinica.edu.tw}  

\maketitle

\abstracts{We study the feasibility of incorporating
the popular Zee model for neutrino mass in the 
framework of the supersymmetric standard model. 
While a singlet slepton has the right quantum number
to play the role of the Zee charged scalar boson, 
the SUSY framework introduces extra contributions. 
We derive conditions for the Zee contributions to 
dominate, hence retaining the flavor of the Zee 
model.}

\section{Zee Neutrino Mass Model}

An economical way to generate small neutrino masses 
with a phenomenologically favorable texture is 
given by the Zee model\cite{zee,fg,jmst}, which 
generates masses via one-loop diagrams.
The model consists of a charged singlet scalar 
$h^{\mbox{-}}_{\mbox{\tiny Zee}}$, 
the Zee scalar, which couples to lepton doublets 
$\psi_{\!\scriptscriptstyle Lj}$ via the interaction
\begin{equation} \label{zcp}
f^{ij} \left( \psi^\alpha_{\!\scriptscriptstyle Li} {\cal C} 
\psi^\beta_{\!\scriptscriptstyle Lj} \right ) 
\epsilon_{\!\scriptscriptstyle \alpha\beta}\; 
h^{\mbox{-}}_{\mbox{\tiny Zee}}\;\;, 
\end{equation}
where $\alpha,\beta$ are the $SU(2)$ indices, $i,j$ are the
generation indices, ${\cal C}$ is the charge-conjugation 
matrix, and $f^{ij}$ are Yukawa couplings antisymmetric 
in $i$ and $j$. The latter fact is a result of the $SU(2)$
product rule and is central to the favorable texture
obtained.  Another ingredient of the Zee model is an 
extra Higgs doublet (in addition to the one that gives 
masses to charged leptons) that develops a 
vacuum expectation value (VEV) and thus provides a mass 
mixing between the charged Higgs boson and the Zee scalar 
boson. The corresponding coupling, together with the 
$f^{ij}$'s, enforces lepton number violation.

A recent analysis by Frampton and Glashow\cite{fg} 
(see also Ref.\cite{jmst}) showed that the Zee mass 
matrix of the following texture 
\begin{equation} 
\label{zee-m}
 \left( \begin{array}{ccc}
0        & m_{e\mu} & m_{e\tau} \\
m_{e\mu} & 0        & \epsilon  \\
m_{e\tau}& \epsilon & 0 
         \end{array}          \right ) \; ,
\end{equation}
where $\epsilon$ is small compared with $m_{e\mu}$ and 
$m_{e\tau}$, is able to provide a compatible mass pattern 
that explains the atmospheric and solar neutrino data. 
The generic Zee model guarantees the vanishing 
of the diagonal elements, while the suppression of the 
$m_{\mu\tau}$ entry, here denoted by the small parameter 
$\epsilon$, has to be otherwise enforced.  Moreover, 
$m_{e\mu} \sim m_{e\tau}$ is required to give the maximal 
mixing solution for the atmospheric neutrino data.

Setting $\epsilon$ to zero in the above mass matrix gives
the following (zeroth order) result: one linear combination 
of $\nu_\mu$ and $\nu_\tau$ remains massless while the
orthogonal state forms a Dirac pair with $\nu_e$. 
Oscillation between the Dirac pair and the massless
Majorana state could explain the atmospheric neutrino
data. If we restore a nonzero $\epsilon$, or, for that 
matter, put in some other perturbation to the above mass 
matrix instead, we have the first order result: namely, we have an 
additonal pseudo-Dirac splitting between the massive
states that could explain the solar neutrino data. To be 
exact, the original massless state would then also gain
a tiny mass. This generalized Zee mass texture is what we
will aim at in our discussion of supersymmetric version(s)
of the Zee model below. 

\begin{figure}[b]
\vspace*{1.5in}
\includegraphics{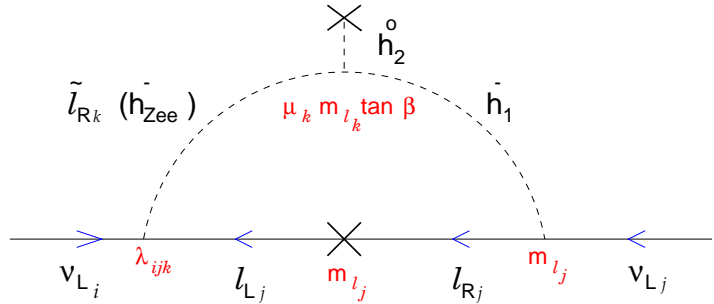}
\caption{SUSY-Zee diagram}
\end{figure}

\section{Zee Model in a Supersymmetric Framework} 
To take the Zee model seriously, one have to put it 
together with other aspects of beyond-standard-model
physics. Hence it worths considering putting the model 
in a supersymmetric framework. The topic is studied in
our recent paper\cite{ck}, upon which the present report
is based.

A naive idea along the line would be to add what is 
needed in the Zee mechanism to the minimal 
supersymmetric standard model (MSSM). However a 
right-handed slepton 
($\tilde{\ell}_{\!\scriptscriptstyle R}$) 
actually has the same gauge
quantum number as the Zee scalar boson, hence the
question of our title --- if the slepton could take the
role. It is clear that, lepton number and, generically 
R-parity, has to be violated here. The MSSM spectrum
even provides the second Higgs doublet needed. A careful study
shows that all we need to complete the SUSY-Zee 
diagram is the following minimal set of only three
R-parity violating (RPV) couplings:
\[ \{\;
 \lambda_{{\scriptscriptstyle 12}\,k}\;,\;
\lambda_{{\scriptscriptstyle 13}\,k}\;,\;
\mu_k \; \}  \nonumber
\]
where family index $k$ can be chosen arbitrarily with
$\tilde{\ell}_{\!\scriptscriptstyle R_k}$ assuming the 
role of the Zee boson. The corresponding diagram is
given in Fig.1. Note that our RPV parameters are defined
in a flavor basis where the so-called sneutrino VEV's
are rotated away (see Refs.\cite{svp,otto} for more 
details).

It is interesting to note that the SUSY-Zee contribution
to neutrino masses involves RPV parameters of both the
bilinear and trilinear type. The existence of such a
contribution under SUSY without R-parity has not been
realized before our work. In fact, there is another  
contribution of the kind that exists even within the 
present minimal framework and, in that case, contributes to the same
neutrino mass entries ($m_{e\mu}$ and $m_{e\tau}$) also 
first identified in Ref.\cite{ck}. This is shown in Fig.2.  

\begin{figure}[b]
\vspace*{1.5in}
\includegraphics{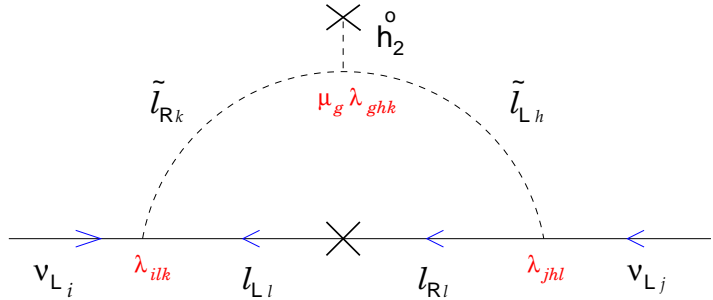}
\caption{Another new contribution with both $\mu_i$ and $\lambda$'s}
\end{figure}

However, the above minimal set of RPV couplings also gives
rise to other contributions that could potentially spoil
the Zee mass texture. These more widely studied
contributions tend to give diagonal mass matrix entries
at least the same strength as the off-diagonal ones,
as illustrated in Ref.\cite{otto}. To retain the 
successful flavor of the Zee model, one has to go to a
region of the parameter space where the SUSY-Zee
contribution dominates hence giving the zeroth order
texture with the sub-dominating contributions fitting 
into a successful final result.

The best scenario here is for $k=3$, {\it i.e.} making
the right-handed stau the Zee boson. We give here the neutrino
mass results and summarize the required conditions as 
follows:
\footnotesize
\beqa
m_{ee} \!&=& \!\!
 \frac{-1}{16\pi^2} \; 
( A^{\!\scriptscriptstyle E}_\tau - \mu \tan\!\beta) 
\;f(M^2_{\tilde{\tau}_L},M^2_{\tilde{\tau}_R}) \;
m_\tau^2\, \lambda_{\scriptscriptstyle 1\!33}^2\;;
\\
m_{\tau\tau} \!&=& \!\!
 \,\frac{- v^2 \cos^2\!\beta \;
( g^2 M_{\scriptscriptstyle 1} + g^{'2} M_{\scriptscriptstyle 2} )}
{2 \mu \;[2 \mu M_{\scriptscriptstyle 1} M_{\scriptscriptstyle 2} 
- v^2 \sin\!\beta \cos\!\beta\; (g^2 M_{\scriptscriptstyle 1} 
+ g^{'2}M_{\scriptscriptstyle 2} ) ]} 
\;\mu_{\scriptscriptstyle 3}^2
\;; \\
m_{e\mu} \!&=& \!\!
 \frac{-1}{16\pi^2} 
\frac{\sqrt{2}\tan\!\beta} {v \cos\!\beta}\;  
f( M_{h_{\scriptscriptstyle 1}^{\mbox{-}}}^2, M_{\tilde{\tau}_R}^2 )
\; m_\tau \, (m_\mu^2 -m_e^2) \,
\mu_{\scriptscriptstyle 3} \lambda_{\scriptscriptstyle 1\!23}
\nonumber \\
&& - \frac{1}{16\pi^2}  \frac{v\sin\!\beta}{\sqrt{2}}\;
 f( M_{\tilde{e}_L}^2, M_{\tilde{\tau}_R}^2 )
\; m_\tau \, \mu_{\scriptscriptstyle 3} \lambda_{\scriptscriptstyle 1\!23}
 \lambda_{\scriptscriptstyle 1\!33}^2
\;; \\
m_{e\tau}\! &=& \!\!
 \frac{-1}{16\pi^2} 
\frac{\sqrt{2}\tan\!\beta} {v \cos\!\beta}\;  
f( M_{h_{\scriptscriptstyle 1}^{\mbox{-}}}^2, M_{\tilde{\tau}_R}^2 )
\; m_\tau \, (m_\tau^2 -m_e^2) \,
\mu_{\scriptscriptstyle 3} \lambda_{\scriptscriptstyle 1\!33} 
\nonumber\\
&& - \frac{1}{16\pi^2}  \frac{v\sin\!\beta}{\sqrt{2}}\;
 f( M_{\tilde{e}_L}^2, M_{\tilde{\tau}_R}^2 )
\; m_\tau \, \mu_{\scriptscriptstyle 3} 
 \lambda_{\scriptscriptstyle 1\!33}^3 
\;;
\eeqa\normalsize
where $f(X,Y) = \frac{1}{X-Y}\;\log\frac{X}{Y}$; and with
$m_{\mu\mu}$ and  $m_{\mu\tau}$ being zero.

The conditions for the success of the scenario are
given by :-
\beqa
\lambda_{\scriptscriptstyle 1\!33} &\sim&\frac{m_\mu^2}{m_\tau^2}\;
\lambda_{\scriptscriptstyle 1\!23}
\;;\\
(\mu_{\scriptscriptstyle 3} \cos\!\beta) \;\lambda_{\scriptscriptstyle 1\!23}
&\sim& \;{\cos^3\!\!\beta}\;{\mbox{Max}( 
M_{h_{\scriptscriptstyle 1}^{\mbox{-}}}^2, 
M_{\tilde{\tau}_R}^2 )}\; (7\cdot10^{-5} \,\mbox{GeV}^{-1})
\;;\\
{\mu_{\scriptscriptstyle 3}^2}\, \cos^2\!\!\beta &\ll& 
{\mu^2} M_{\scriptscriptstyle 1}\,(1\cdot10^{-14}\, {\rm GeV}^{-1}) 
\;;\\
\lambda_{\scriptscriptstyle 1\!33}^2
&\ll& \frac{\mbox{Max}(M^2_{\tilde{\tau}_L},M^2_{\tilde{\tau}_R})}
{(A^{\scriptscriptstyle E}_\tau - \mu \tan\!\beta)}\;
 (2.5 \cdot 10^{-9} \,\mbox{GeV}^{-1}) \;;\\
\frac{\lambda_{\scriptscriptstyle 1\!23}^2}{M^2_{\tilde{\tau}_R}} &\leq&
10^{-8}\,\mbox{GeV}^{-2} \;. 
\eeqa \normalsize 
The feasibility of the scenario is marginal. Alternative versions with
some extra chiral superfields, introduced and briefly discussed in
Ref.\cite{ck}, are much less contrained though.


\end{document}